\documentclass[aps,prb,floatfix,twocolumn,showpacs,superscriptaddress]{revtex4-1}
\usepackage{amsmath,bm,graphics,epsfig,amssymb}
\usepackage{mathrsfs}
\usepackage{color}
\usepackage{CJK}
\usepackage{tikz}
\usepackage[latin1]{inputenc}
\usepackage[normalem]{ulem}

\newcommand{\mac}{\mathcal}

\newcommand{\tx}{\text}
\newcommand{\ti}{\textit}
\newcommand{\tb}{\textbf}

\newcommand{\la}{\langle}

\newcommand{\nn}{\nonumber}

\newcommand{\pr}{\prime}
\newcommand{\ra}{\rangle}
\newcommand{\raw}{\rightarrow}
\newcommand{\til}{\tilde}

\newcommand{\alp}{\alpha}
\newcommand{\dlt}{\delta}

\newcommand{\lam}{\lambda}

\newcommand{\og}{\omega}

\newcommand{\sg}{\sigma}

\newcommand{\ie}{{i.e.,~}}

\newcommand{\BP}{\hat{\bm{\sigma}}}
\newcommand{\pmx}{\hat{\sigma}_{x}}
\newcommand{\pmy}{\hat{\sigma}_{y}}

\newcommand{\bk}{\bm{k}}

\newcommand{\br}{\bm{r}}
\newcommand{\EE}{\bm{E}}
\newcommand{\sgn}[1]{{\rm{sgn}(#1)}}

\newcommand\rmv{\bgroup\markoverwith {\textcolor{red}{\rule[0.5ex]{2pt}{0.4pt}}}\ULon}

\newcommand{\Eq}[1]{Eq.~\eqref{#1}}

\newcommand{\Figure}[1]{Fig.~\ref{#1}}

\begin{document}
\begin{CJK*}{UTF8}{gbsn} 

\title{Spin-Hall Conductivity and Electric Polarization in Metallic Thin Films}
\author{Xuhui Wang}
\affiliation{King Abdullah University of Science and Technology
(KAUST), Physical Science and Engineering Division, Thuwal
23955-6900, Saudi Arabia}
\author{Jiang Xiao (萧江)}
\email[Corresponding author:~]{xiaojiang@fudan.edu.cn}
\affiliation{Department of Physics and State Key Laboratory of
Surface Physics, Fudan University, Shanghai 200433, China}
\affiliation{Center for Spintronic Devices and Applications, Fudan University, Shanghai 200433, China}
\author{Aurelien Manchon}
\affiliation{King Abdullah University of Science and Technology (KAUST),
Physical Science and Engineering Division, Thuwal 23955-6900, Saudi Arabia}
\author{Sadamichi Maekawa}
\affiliation{Advanced Science Research Center,
Japan Atomic Energy Agency, Tokai 319-1195, Japan}
\affiliation{CREST, Japan Science and Technology Agency, Tokyo 102-0075, Japan}
\date{\today}

\begin{abstract}
We predict theoretically that, when a normal  metallic thin film
(without \ti{bulk} spin-orbit coupling, such as Cu or Al) is
sandwiched by two insulators, two prominent effects arise due to
the {\it interfacial} spin-orbit coupling: a giant spin-Hall
conductivity due to the surface scattering and a transverse
\ti{electric} polarization due to the spin-dependent phase shift
in the spinor wave functions.
\end{abstract}
\maketitle
\end{CJK*}

Spin-orbit interaction, transferring angular momentum between
electronic spins and orbital motion, has extended the boundary of
the field of spintronics towards a full-electric manipulation of
spins without using magnets. By coupling the charge and spin
currents, spin-orbit interaction has left its signature in bulk
metals \cite{hirsch-she,zhang-she,taka-mae-she} and semiconductors
\cite{murakami-science,sinova-prl-she-2004,kato-science-she-2004,bauer-science,
wunderlich-prl-2005} by the so-called spin-Hall effect,
\cite{dp-she-1971} which catches much attention in academia and
industry due to its interesting physics and potential
applications.

In noble metals such as Pt and Au, a strong bulk spin-orbit
coupling drives the (inverse) spin-Hall effect, which has been
thoroughly studied in numerous settings, providing an alternative
to generate or detect spin current.
\cite{valenzuela-2006,spin-hall-kimura,spin-hall-ando,spin-hall-mosendz}
Theoretical investigations, \ti{ab initio} calculations in
particular, have strengthened our understanding of the role played
by bulk impurities \cite{gradhand-prl,lowitzer-prl} and doped
surfaces. \cite{gu-prl} With a series of efforts to dope Cu using
spin-orbit scatterers (see Ref.\onlinecite{niimi-prl} and
references therein), a large spin-Hall angle has been reported in
Cu thin film with bismuth impurities recently. \cite{niimi-2012}
In this Rapid Communication, we propose an alternative method to
realize the spin-Hall effect in Cu without doping but by
sandwiching the Cu thin film with two dissimilar insulators such
as oxides or even vacua. The inversion symmetry breaking across
the interfaces provides interfacial spin-orbit couplings (ISOCs),
thus allowing metals such as Al or Cu to accommodate a spin-Hall
conductivity that may be even larger than that caused by a bulk
spin-orbit interaction in noble metals. Meanwhile, we demonstrate
that the ISOC also induces an in-plane wave-vector-dependent
spatial separation of wave functions between opposite spin types
along the confinement direction. Combined with the spin imbalance
(between the majority and minority bands) due to the structural
asymmetry, such a spatial separation induces a transverse electric
polarization that is quadratic in the in-plane electric field.

\begin{figure}[b]
\centering
\includegraphics[width=.48\textwidth]{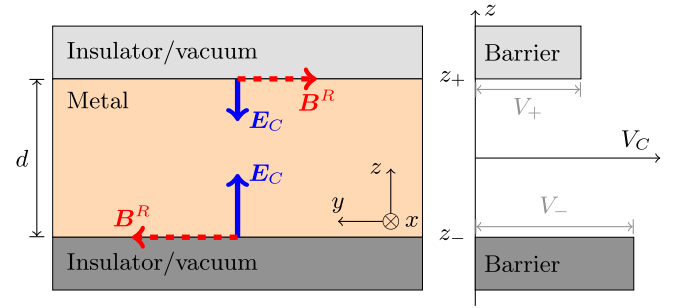}
\caption{(Color online) Left: A metallic film of thickness $d$ is
sandwiched by two dissimilar insulating materials, such as oxides
or vacua. The electric fields (solid blue arrows) at the
interfaces point inward (${\pm}z$), and the resulting Rashba
magnetic fields (dashed red arrows) point sideward (${\pm}y$) for
an electron moving along the $x$ direction. Right: The confining
potential $V_C$ in \Eq{eq:hami-confine} along the confinement
direction (\ie $z$ axis).} \label{fig:setup}
\end{figure}

A schematic picture of the setup is shown in \Figure{fig:setup}.
The normal metal film has a constant thickness $d$ along the
transverse direction ($z$), while two interfaces are in the $xy$
plane located at $z_{\pm}={\pm}d/2$. The confinement of the
insulators is described by finite potential steps,
\begin{equation}
    V_{C}=V_{+}{\theta}(z-z_+) + V_{-}{\theta}(z_{-}-z),
\label{eq:hami-confine}
\end{equation}
where $V_{\pm}$ is the height of the potential barrier  at
$z_{\pm} = {\pm}d/2$ and ${\theta}(z)$ is the Heaviside step
function.  We are interested in the surface scattering by the
Rashba-type spin-orbit interaction \cite{rashba-1984} generated at
interfaces between a normal metal film and insulating materials,
such as metal oxides.  The potential barriers \Eq{eq:hami-confine}
generate two electric fields $\bm{E}_C = -\bm{\nabla}V_C$ that are
localized at the interfaces and aligned oppositely to each other
along the $z$ direction, giving rise to a Rashba-type ISOC,
\begin{equation}
    H_{R}=
    {\lam_-V_-\dlt(z-z_-)-\lam_+V_+\dlt(z-z_+)\over {\hbar}}
    \left(\pmx\hat{p}_{y}-\pmy\hat{p}_{x}\right),
\label{eq:hami-soi}
\end{equation}
where $\lam_{\pm}$ is the spin-orbit coupling parameter for the
corresponding interface, $\BP$ is the Pauli matrix, and
$\hat{\bm{p}}$ is the canonical momentum.

\begin{figure}[t]
    \centering
    \includegraphics[width=.49\textwidth]{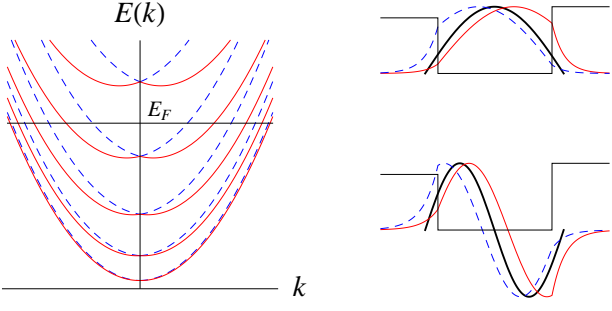}
    \caption{(Color online) Left: Schematic view of the band
    structure in metallic thin film with interfacial Rashba spin-orbit coupling,
    majority (solid red) and minority (dashed blue).
    Right: Schematic picture of the spatial part of the
    wave-function for the first two subbands; the stepwise
    curve is the confinement potential profile and the three different
    curves are the wave function from the lowest-order perturbation
    \Eq{eq:wave-func-dp} (solid black for both majority and
    minority) and
    the exact wave function from \Eq{eq:exact_wf} (solid red for majority
    and dashed blue for minority) which shows the spatial separation of
    opposite spins. The discontinuity at the interfaces is due to the
    ${\delta}$-function-like potential in $H_R$. }
\label{fig:band_wave}
\end{figure}

\ti{Band structure and wave function.} The full Hamiltonian for
this system is $H = {\bm p}^2/(2m) + V_C + H_R$.  We first treat
the ISOC $H_{R}$ with degenerate perturbation.  Using the wave
functions of a finite potential well and a standard degenerate
perturbation technique, \cite{landau-lifshitz-qm} assuming the
energy splitting due to the ISOC is much smaller than the
interchannel energy spacing, we obtain the energy eigenvalue and
wave function for an eigenspinor labeled by spin polarization
$s={\pm}$, an in-plane wave vector $\bk$, and a transverse channel
index $n$,
\begin{subequations}
\label{eq:band-wave}
\begin{align}
    E_{n\bk s}&=\frac{{\hbar}^{2}k^{2}}{2m}+E_{0}n^{2}
    +s\frac{2}{d_{e}}k|\lam_{+}-\lam_{-}|E_{0}n^{2},
    \label{eq:energy-dp} \\
    {\psi}_{n\bk s}&=\frac{1}{\sqrt{A d_{e}}}e^{i\bk{\cdot}\br}
    \sin\left(\frac{n{\pi}}{d_{e}}z+\dlt\right)
    \left(\begin{array}{c}
    1 \\ -i s e^{i{\theta}}
\end{array}\right),
\label{eq:wave-func-dp}
\end{align}
\end{subequations}
where $\tan{\theta}= k_y/k_x$,
$E_{0}{\equiv}{\hbar}^{2}{\pi}^{2}/(2 m d_{e}^{2})$, and $A$ is
the area of the metal film. A schematic view of the band energy
and the spatial part of the wave function is shown in
\Figure{fig:band_wave}. The spatial part of the wave function
${\psi}_{n\bk s}$ in \Eq{eq:wave-func-dp} is the same for majority
($s = -$) and minority ($s = +$) spins in the first-order
perturbation calculation. The phase shift for both spins is given
by $\dlt=n{\pi}/2+(d_+-d_-)n{\pi}/(2d)$ with
$d_{\pm}={\hbar}/\sqrt{2mV_{\pm}}$ the approximate penetration
depth into the barrier.  By assuming high confinement barriers
($V_{\pm}{\gg}E_F$, given $E_{F}$ the Fermi energy), the effective
thickness $d_{e}=d+d_++d_-$.  Therefore, a finite potential well
of thickness $d$ can be viewed equivalently as an infinite well
with thickness $d_{e}$. We shall point out that, although the
eigenenergies for majority and minority are different, such a
difference does not lead to any magnetism because of the exact
cancellation among spins at different $\bk$ directions.

\ti{Spin-Hall conductivity.} When $\lam_{+}{\neq}\lam_{-}$, for
each conducting channel (or transverse mode), the electron is
characterized by a free Hamiltonian
$H_{0}^{(n)}=\bm{p}^{2}/(2m)+E_{0}n^{2}$ augmented by an effective
momentum-dependent Rashba magnetic field
$\bm{B}^{\rm{R}}_n(\bm{p}) = (2{\lam}^{\rm
eff}_n/{\hbar})(\hat{\bm{z}}{\times}\bm{p})$ with a
channel-dependent coupling constant ${\lam}^{\rm eff}_n =
2({\lam}_{+}-{\lam}_{-})E_0 n^2/d_e$.  For an electron in band $n$
carrying an in-plane momentum $\bm{p}$, assuming ${\lambda}_+ >
{\lambda}_-$, the majority (minority) spin points to the
$\bm{B}_{n}^{\rm{R}}$ ($-\bm{B}^{\rm{R}}_{n}$) direction.

When an electric field is applied along the $\hat{x}$ axis, the
Rashba field $\bm{B}^{\rm{R}}_n(\bm{p}(t))$ becomes time dependent
since $\dot{p}_x(t) = -eE_x$. \cite{sinova-prl-she-2004} In the
rotating frame that follows $\bm{B}^{\rm{R}}_n(\bm{p}(t))$, the
time dependence is translated into a gauge field
$\bm{B}^{\rm{G}}_n = -e{\hbar}p_y E_x/p^2\hat{\bm{z}}$.
\cite{aha-stern}  In the adiabatic limit
($eE_{x}{\ll}\lam_{n}^{\rm{eff}}k_{F}^{2}$), the majority
(minority) spins align (antialign) with the total field
$\bm{B}^{\rm{T}}_n = \bm{B}^{\rm{R}}_n + \bm{B}^{\rm{G}}_n$,
resulting in an out-of-plane ($\hat{z}$) spin component
\begin{equation}
    n_z^{s}(\bm{p}) {\approx} s{B^{\rm{G}}_n\over B^{\rm{R}}_n}
    = -s{e{\hbar}^2p_yE_x\over 2 {\lambda}_{n}^{\rm eff} p^3}.
\label{eqn:nz}
\end{equation}
Consequently, in a metal film of volume $\mac{V}=d A$, at the
Fermi level, the spin current polarized in the $\hat{z}$ direction
while flowing in the $\hat{y}$ direction is
\begin{equation}
    j_{y}^{z} = \frac{1}{\mac{V}}{\sum}_{n,\bm{p},s}
    {{\hbar}\over 2}n_z^s(\bm{p})\frac{p_y}{m}
    = -\frac{e n_{c}}{8{\pi}d}E_{x}.
\label{eqn:jsy}
\end{equation}
Therefore, the spin-Hall conductivity caused by ISOC is
\begin{equation}
    {\sigma}^{\rm{SH}} = \frac{e}{8{\pi}}\frac{n_{c}}{d}
    {\approx} {e\over 8{\pi}^2}k_F.
\label{eqn:sigma-sH}
\end{equation}
where $n_{c}=\lfloor k_{F}d/{\pi}\rfloor$ is the total number of
transverse channels. The symbol $\lfloor a \rfloor$ denotes the
largest integer that is smaller than $a$. Note that
\Eq{eqn:sigma-sH} is obtained for $\lam_{+}>\lam_{-}$, the sign of
${\sigma}^{\rm SH}$ should change when $\lam_+<\lam_-$, and
${\sigma}^{\rm SH} = 0$ for $\lam_{+}=\lam_{-}$.

To extend the semiclassical picture above, we employ the Kubo
formula to calculate the dc spin-Hall conductivity (\ie $\og\raw
0$) in the linear response regime: \cite{sinova-prl-she-2004} An
electric current flowing along the $x$ direction gives rise to a
spin current that is polarized to the $z$ axis while transporting
along the $y$ direction,
\begin{align}
    \sg^{\rm{SH}}
  = & \frac{e{\hbar}}{V} {\sum}_{\bk}{\sum}_{n,n^{\pr}}{\sum}_{s{\neq}s^{\pr}}
    (f_{n'\bk s'}-f_{n\bk s})\nn\\
     &{\times}\frac{ \tx{Im}\left[\la {\psi}_{n'\bk s'}|\hat{j}_{y}^{z}|{\psi}_{n\bk s}\ra
    \la {\psi}_{n\bk s}|\hat{v}_{x}|{\psi}_{n'\bk s'}\ra\right]}
     {(E_{n\bk s}-E_{n'\bk s'})(E_{n\bk s}-E_{n'\bk s'}-{\hbar}\og-i{\eta})},\nn\\
\label{eq:spin-hall-cond}
\end{align}
where $f_{n\bk s}$ is the electron spin occupation number. Using
the Heisenberg equation, the in-plane velocity operator is given
by
\begin{equation}
    \hat{v}_{x(y)} = \frac{\hbar}{i m}{\nabla}_{x(y)}
    {\mp}\frac{\lam_-V_-\dlt(z-z_-)-\lam_+V_+\dlt(z-z_+)}{\hbar}\hat{\sg}_{y(x)},
\label{eq:anomalous-velocity}
\end{equation}
where the second term is the anomalous velocity due to the
spin-orbit coupling that is localized at the interface. The
definition of spin current polarized along the $z$ direction is
$\hat{\bm{j}}^{z}=({\hbar}/4)\{\hat{\sg}_{z},\hat{\bm{v}}\}$.
Using the eigensolutions in \Eq{eq:band-wave}, we have
\begin{equation}
    \sg^{\rm{SH}}
    = \frac{e}{8{\pi}} \frac{\lam_+-\lam_-}{|\lam_+-\lam_-|} \frac{n_{c}}{d}
    ={\pm}\frac{e}{8{\pi}}\frac{n_{c}}{d}
    {\approx}{\pm}\frac{e}{8{\pi}^{2}} k_{F},
\label{eq:spin-hall-general-kubo}
\end{equation}
where ${\pm}= \sgn{\lam_+-\lam_-}$ and the last approximate value
assumes a large number of transverse channels $n_{c}{\gg} 1$. For
two perfectly identical interfaces ${\lambda}_{-}={\lambda}_{+}$,
the spin-Hall conductivity shall vanish. Equation
(\ref{eq:spin-hall-general-kubo}) calculated from Kubo formula
agrees with \Eq{eqn:sigma-sH} derived semiclassically and
comprises one of the main results of this communication.

To describe a spin-Hall system, one quantity often scrutinized in
experiments is the spin-Hall angle defined as the ratio between
$\sg^{\rm{SH}}$ and the longitudinal conductivity $\sg^{\rm{N}}$,
characterizing the efficiency of converting charge (spin) current
into spin (charge) current in a (inverse) spin-Hall system. For a
coherent ballistic conductor presented in this communication, the
dc longitudinal conductance (or resistance) measured in an
experiment is dominated by scattering events at contacts.
\cite{bauer-prb-2003} Therefore, we expect the spin-Hall angle to
depend on the specific geometry and material selection of the
contacts.

\ti{Transverse electric polarization.} We now turn to an inviting
effect, as caused by ISOC, that has not been discussed before to
our knowledge. The ISOC in \Eq{eq:hami-soi} not only gives rise to
an energy splitting for opposite spins, but also separates the
wave functions in real space; \ie majority and minority spins are
shifted towards opposite surfaces. This can be easily understood
as the following: Assuming $\lam_+>\lam_-$, the potential barrier
at surface $z_+$ ($z_-$) is decreased (increased) by a
$\dlt$-function Rashba potential for majority spins; therefore
majority spins tend to shift towards surface $z_+$. On the
opposite, the minority spins shift toward surface $z_-$.
Therefore, majority and minority spins are spatially separated
along the transverse direction as shown schematically in
\Figure{fig:band_wave}.  To quantify such an effect, the
approximated wave function in \Eq{eq:wave-func-dp} is not enough,
and we must seek an exact eigensolution to the full Hamiltonian
$H$ of the following form (inside the metal film):
\begin{equation}
    {\psi}_{n\bk s} = c_{n\bk s} e^{i\bm{k}{\cdot}\bm{r}}
    \sin(q_{n\bk s} z+{\delta}_{n\bk s})
    \left(\begin{array}{c} 1 \\ -i s e^{i{\theta}} \end{array}\right),
\label{eq:exact_wf}
\end{equation}
with normalization factor $c_{n\bk s}$ and the expression outside
the metal is written similarly but with evanescent wave function
in the $z$ direction. In the limit of ${\lambda}_{\pm}k_F^2{\ll}
1$ and $E_F{\ll}V_{\pm}$,
\begin{subequations}
\label{eq:spin-dep-phase-shift}
\begin{align}
q_{n\bk s} &{\approx}\frac{n{\pi}}{d}+\frac{n{\pi}}{d^{2}}
\left[sk(\lam_+-\lam_-)-\left(d_+ + d_-\right)\right],\\
\dlt_{n\bk s} &{\approx}\frac{n{\pi}}{2}-\frac{n{\pi}}{2d}
\left[sk(\lam_++\lam_-)-\left(d_+ - d_-\right)\right].
\end{align}
\end{subequations}
The spin-dependent transverse wave vector $q_{n\bm{k}s}$ gives the
energy splitting  as in \Eq{eq:energy-dp} for the majority and
minority spins with $E_{n\bk s} = ({\hbar}^2/2m)(q_{n\bk s}^2 +
k^2)$. The spin-dependent phase shift $\dlt_{n\bk s}$ means that
the wave function for majority (minority) spins with $s = -(+)$ is
shifted toward negative (positive) $z$ direction. The transverse
shift is spin and $k$ dependent, and can be quantified by the
center of probability in the $z$ direction:
\begin{equation}
z_{ns}(\bk) = {\int}z |{\psi}_{\bm{k}ns}(z)|^2 dz
{\approx}sk{{\lambda}_{+} + {\lambda}_{-}\over 2}.
\label{eq:zshift}
\end{equation}
The last approximation omits a constant $(d_{+}-d_{-})/2$ that
does not contribute to the nonequilibrium properties discussed
below. For highly asymmetric interfaces (\ie
${\lambda}_+{\gg}{\lambda}_-$), the transverse shift
$|z_{n+}(\bk)-z_{n-}(\bk)|{\propto}{\eta}(k) d$, where ${\eta}(k)$
is the ratio between the interfacial Rashba energy splitting
[third term in \Eq{eq:energy-dp}] and transverse channel energy
spacing [second term in \Eq{eq:energy-dp}]. Therefore, by a
realization of large ${\eta}$, the majority and minority spin
channels can be spatially separated, and the spin-flip scattering
is thus suppressed.

We now discuss the effect of the transverse shift on the electric
polarization. What interests us the most is the nonequilibrium
response of the electric polarization as a direct consequence of
the ISOC.  Because of the $k$ dependence, the transverse shift
$z_{ns}(\bk)$ depends on the application of a current; therefore a
nonequilibrium transverse electric polarization response to the
in-plane electric field $\EE$ can be calculated as $P_z(\bm{E})=
\frac{e}{\mac{V}}{\sum}_{n\bm{k}s}[z_{ns}(\bk+{\delta}\bk)-
z_{ns}(\bk)]$ with ${\delta}\bm{k} = e{\tau}\bm{E}/{\hbar}$:
\begin{subequations}
\label{eq:electric-polarization}
\begin{align}
P_z(\bm{E})
&= -\frac{ek_F^3}{3d}\left(\frac{e{\tau}}{h}\right)^{2}
(\lam_+^2 -\lam_-^2)\bm{E}{\cdot}\bm{E}
\label{eq:electric-polarization-field}\\
&=-\frac{m^{2}e^{3}}{12{\pi}^{2}{\hbar}^{4}}(\lam_{+}^{2}-\lam_{-}^{2})
\frac{k_{F}}{d}V^{2}.
\label{eq:electric-polarization-volt}
\end{align}
\end{subequations}
The second expression is for a coherent conductor with an electron
dwelling time ${\tau}=l/v_{F}$ and $V = El$. The quadratic
dependence of $\bm{E}$ or $V$ is a result of symmetry in the
in-plane dimensions.

In an analogy to Hall effects, the electric polarization
\Eq{eq:electric-polarization} shall give rise to, across the
confinement direction, a voltage signal that is \ti{quadratic} in
the (in-plane) applied voltage. The electric polarization
\Eq{eq:electric-polarization} is a combination of two facts: (1)
the spatial separation between majority and minority spins, which
is proportional to $\lam_++\lam_-$, and (2) the spin imbalance
between the majority and minority spins, which is proportional to
$\lam_+-\lam_-$. Equation (\ref{eq:electric-polarization})
comprises the other main result of this communication.

We provide an estimate on the order of magnitude of the
polarization density for the most asymmetric interfaces (e.g.,
$\lam_{+}{\neq}0$ and $\lam_{-}= 0$). The existing value of Rashba
parameter in literature ranges from $\alp_{R}=0.16$~\tx{eV}~\AA
~\cite{soi-value-prb} to
$\alp_{R}=2.5$~\tx{eV}~\AA,~\cite{soi-value-prl} depending on the
materials. Here, we take $\alp_{R}=0.16$~\tx{eV}~\AA,
corresponding to a Rashba energy splitting
${\Delta}_{R}=0.22$~\tx{eV}, for Cu with $k_{F}=1.36$~\AA$^{-1}$.
We convert such an energy scale into $\lam$ by
${\Delta}_{R}{\approx}4k_{F}\lam E_{0}/d$. For a thin film of a
thickness $d=10$~nm and under an applied voltage, for example
$V=100$~nV, the polarization density
$P_{z}{\approx}2.1{\times}10^{-14}$~C~m$^{-2}$. Such an electric
polarization is measurable (in a circuit enclosing two interfaces)
under an ac in-plane electric field (or current): With a frequency
$f = 1~\tx{GHz}$, the magnitude of the induced current density  in
the transverse direction is $j_z {\sim} dP_z/dt {\sim}
10^{-5}$~A~m$^{-2}$. Meanwhile, the frequency of this induced
current is doubled to $2f$ because $P_z$ depends quadratically on
$V$. Such a frequency doubling, as a consequence of the symmetry
of the present system, is qualitatively different from Pershin and
Di Ventra in Ref. \onlinecite{pershin-diventra-2009}, where the
frequency doubling effect emerges from electron-electron
interaction. In addition, we emphasize the electric polarization
in our study arises from the phase shift of a single-electron wave
function along the out-of-plane $z$ direction, while in Ref.
\onlinecite{pershin-diventra-2009}, the electric polarization
arises from a nonlinear effect due to many-body electron-electron
interaction and an inhomogeneous charge density.

To the best of our knowledge, the transverse electric polarization
induced by the ISOC predicted in this communication is a new
effect that has not been discussed previously. This electric
polarization manifests itself as a {\it charge} Hall effect and
can be measured as a transverse voltage (or a current in a close
circuit). Being qualitatively different from conventional charge
Hall effects driven by Lorentz force in normal metals and
anomalous Hall effect in ferromagnetic media, the electric
polarization [Eq.(\ref{eq:electric-polarization})] is quadratic
(instead of linear) in the longitudinal electric field and does
not require external magnetic field or ferromagnetism.

\ti{Discussion and Conclusion.} The spin-Hall conductivity in
\Eq{eq:spin-hall-general-kubo} is derived for a ballistic sample
where the bulk impurities are scarce. This is generally valid for
thin film with thickness less than the mean free path and spin
diffusion length. In an ultrathin film, surface roughness is a
dominating scattering mechanism. \cite{tjm-prl-1986} In the
present setting, the electric fields generated by the potential
gradients are normal to interface while the presence of surface
roughness effectively randomizes the field around an average
direction that is still perpendicular to the average interface.
Therefore, we can summarize the total effect of surface roughness
into an effective spin-orbit coupling $\til{\lam}_{i}$ that is
smaller than $\lam_{i}$. Since the leading order spin-Hall
conductivity does not depend on $\lam$'s, we argue that the
spin-Hall effect survives the roughness scattering. The seemingly
abrupt jump in spin-Hall conductivity [\ie
\Eq{eq:spin-hall-general-kubo}] at the point $\lam_{+}=\lam_{-}$
is the manifestation of neglecting electron momentum relaxation in
such a ballistic conductor.  In a realistic setup, the
sandwich-type conductor is always connected to reservoirs where
the electrons are relaxed and a smoother change shall appear.

We need to point out that these results are, in many ways,
different from the seminal works by Sinova \ti{et al.} in a
two-dimensional electron gas \cite{sinova-prl-she-2004} and
Murakami \ti{et al.} in a bulk semiconductor.
\cite{murakami-science}  First of all, in the present setup the
spin-orbit coupling is neither \ti{intrinsic} to the electrons (as
in Ref. \onlinecite{sinova-prl-she-2004}) nor arising from a
particular band structure (as in Ref.
\onlinecite{murakami-science}), but due to the interface
scattering. Second, our treatment to the spin-Hall conductivity is
seemingly two-dimensional but the outcome highlights a bulk effect
with a weak thickness ($d$) dependence, as long as the structure
can be treated coherently.

Third, and most importantly, because of the finite size in the $z$
direction, such a sandwich-type structure also accommodates a
transverse electric polarization (being qualitatively different
from the spin polarization studied by Edelstein
\cite{edelstein-1990}) through the ISOC, which is unique for the
thin-film structure: The \ti{out-of-plane} electric polarization
along the $z$ direction does not exist in two-dimensional systems
considered, for example, in Refs.\onlinecite{sinova-prl-she-2004,
pershin-diventra-2009, edelstein-1990}.

In conclusion, we predict that in a coherent ballistic conductor
that consists of an ultrathin normal metal (with negligible bulk
spin-orbit coupling) film sandwiched by dissimilar insulators, the
Rashba-type ISOC supports a spin-Hall effect featured by a large
spin-Hall conductivity that is independent of the interfacial
Rashba coupling. The ISOC also causes a spatial separation in the
transverse wave functions of different spin bands. In response to
an in-plane current or electric field, such a spatial separation
gives rise to a transverse electric polarization that is quadratic
in the in-plane field applied. The sandwich-type structure
proposed in this communication has potential applications to
replace noble metals (such as Pt) as a source and detector for
spin currents.

We thank Gerrit E. W. Bauer, Xiaofeng Jin and Mark Stiles for
stimulating discussions. This work was supported by the National
Natural Science Foundation of China (Grants No. 11004036, and No.
91121002) and  the special funds for the Major State Basic
Research Project of China (No. 2011CB925601).


\end{document}